\documentstyle[aps,titlepage,psfig]{revtex}
\begin{document}
\title{One-particle spectral function of electrons in a hot and dense plasma}
\author{A. Wierling and G. R\"opke}
\address{Fachbereich Physik, Universit\"at Rostock\\
         Universit\"atsplatz 1\\
         D 18051 Rostock, Germany}
\date{\today}
\maketitle
\begin{abstract}
A self-consistent determination of the spectral function and the
self-energy of electrons in a hot and dense plasma is reported.
The self-energy is determined within the approximation of the screened
potential. It is shown, that the quasi-particle concept is not an
adequate concept for hot and dense plasmas, since the width of the
spectral function has to be considered. As an example, the solar core
plasma is discussed. An effective quasi-particle picture is
introduced
and results for the solar core plasma as well as for ICF plasmas are
presented.
\end{abstract}
%%%%%%%%%%%%%%%%%%%%%%%%%%%%%%%%%%%%%%%%%%%%%%%%%%%%%%%%%%%%%%%%%%%%%%%%
\section{Introduction}
\label{intro}
Dense plasmas are intensively studied both in experimental as well as
theoretical physics.  Typical plasmas of this kind can be found in
astrophysical objects like the interior of the sun and the giant
planets.  On earth, very dense and hot plasmas are investigated in the
context of inertial confinement fusion. Furthermore, the electron gas
in some metals and semiconductors represents a dense and cold plasma.
Form the study of dense plasmas, important information about
microscopic mechanisms like the interplay between collective effects
and collisions can be gained.  The consistent description of this
interplay is a very challenging task. 

In a dense plasma, the propagation of a single-particle excitation is
a collective phenomenon which is characterized by the single-particle
spectral function. The spectral function for the specie $a$ is given as
the discontinuity of the single-particle Green's function at real
frequencies
\begin{eqnarray}
  A_a(p,\omega) & = & i\hbar \left( G_a(p,\omega+i0) -
    G_a(p,\omega-i0)\right) \;\;\;.
\end{eqnarray}
In contrast to the quasi-particle picture, where the spectral function
is assumed to be proportional to the $\delta$-function,
\begin{eqnarray}
  \label{quasi}
  A_a(p,\omega) & = & 2\,\pi\,\delta\left(\hbar \omega-E_a(p)\right)\;\;\;,
\end{eqnarray}
in dense plasmas 
the broadening of the spectral function is of 
great importance, since it describes the finite life time of the
excitations in the plasma. Here $E_a(p)$ denotes the quasi-particle
energy defined below.
The spectral function has a number of interesting properties, which
are connected to the thermodynamical properties of the plasma. 
First of all, there is a frequency sum rule
\begin{eqnarray}
  \int_{-\infty}^{\infty}\,d\omega\,
A_a(p,\omega) & = & 2\,\pi\;\;\;.
\end{eqnarray}
As a consequence, the spectral function gives the probability to find
a certain frequency at a given momentum $p$.
Furthermore, the density of states is related to the spectral function
by
\begin{eqnarray}
  D_a(\omega) & = & \int\!\frac{d^3p}{(2\,\pi)^3}\,A_a(p,\omega)\;\;\;,
\end{eqnarray}
leading to the so called density relation
\begin{eqnarray}
  \label{density_relation}
  n_a(\mu_a,\beta) & = & \int_{-\infty}^{\infty}\frac{d\hbar
    \omega}{2\,\pi}
 f_a(\omega)\,D_a(\omega)\;\;\;\;,
\end{eqnarray}
where $f_a$ denotes the Fermi distribution function. For a system with
a given density $n_a$, this relation can be used to fix the
corresponding chemical potential $\mu_a$. Additional thermodynamic
properties can be derived, e.g. the equation of state can be found.
Therefore, the equation of state can be improved taking into account
many-particle effects via an appropriate spectral function.

The spectral function for dense systems has been determined in the
context of nuclear physics as well as solid state physics. In nuclear
physics \cite{ars96}, the spectral function of nuclear matter has been
studied extensively. It has been shown that the spectral function
exhibits a complex energy dependence, which cannot incorporated in a
simple quasi-particle picture. Effects of higher order correlations
like the pairing instability on the spectral function have been found.
The influence of these pairing effects are reduced if the 
broadening of the spectral function increases.

In solid state physics, extensive studies of the spectral function
have been carried out within the Hubbard-model \cite{dhs96} and the
t-J model\cite{dag94}. Thermally broadened quasi-particles have been
found at high temperatures. Zimmermann et al. \cite{zs86} performed
a expansion with respect to the width of the spectral function to get
a so called extended quasi-particle picture, but its use is limited to
small deviations from the quasi-particle regime. A calculation similar
to the one reported here, has been carried out by Barth and Holm
\cite{bh96}. They applied their calculation to the electron gas at
zero temperature.  In solid state physics the spectral function can be
measured using angle-resolved photoemission spectroscopy.

\section{ Spectral function and self-energy} 

The single-particle Green's function of the specie $a$ reads
\begin{eqnarray}
  G_a(p,z) & = & \left( \hbar z -\frac{\hbar^2 p^2}{
 2\,m_a}-\Sigma_a(p,z) \right)^{-1}
  \;\;\;\;.
\end{eqnarray}
The medium modifications enter via the self-energy $\Sigma_a$, 
\begin{eqnarray}
  \int\!d\bar 1\,\Sigma_a(1\bar 1)\,G_a(\bar 1 1') & = &
  - \sum_b \int\!d2\, V(1-2) \,G_{ab}(121'2^+)
\end{eqnarray}
where higher order correlations are hidden in the two particle
Green's function $G_{ab}$.
The numbers label the position and time variables and the potential
$V$ denotes the Coulomb-Potential. 
The spectral function can be related to the self-energy 
$\Sigma_a(p,\omega)$ via Dyson's equation
\begin{eqnarray}
  \label{dyson}
    A_a(p,\omega) & = &
  \frac{2\,\mbox{Im}\,\Sigma_a(p,\omega)}{\left(\hbar\,\omega-
  \frac{\hbar^2\,p^2}{2\,m_a}
  -\mbox{Re}\,\Sigma_a(p,\omega)\right)^2\,+\,
  \left( \mbox{Im}\,\Sigma_a(p,\omega) \right)^2}\;\;\;.
\end{eqnarray}
This relation shows, that the imaginary part of the self-energy plays
the role of a width of the spectral function while the real part acts
as the shift of the free dispersion relation. Different approximations
for the self-energy apply for different systems taking into account 
different collective mechanisms. For high density systems interacting 
via the Coulomb potential, Hedin \cite{hed65} proposed the so called GW
approximation where polarization effects are considered. In nuclear
matter, particle-particle correlations are the leading mechanisms 
as was pointed out by Galitski \cite{gal58}. Therefore, a T-matrix
approximation of the self-energy has to be used. A consistent
determination of the spectral function using this approximation for
the self-energy was performed by Alm et al. \cite{ars96}.

Within the GW approximation, the correlated part
of the self-energy is determined by
\begin{eqnarray}
  \label{hedin}
  \Sigma_a^{corr}(p,z) & = &
  -\!\int_{-\infty}^{\infty}\!\frac{d\omega'd\omega}{(2\pi)^2}\!
  \int\!\frac{d^3q}{(2\pi)^3}
  V(q)\frac{2\,\mbox{Im}\,\epsilon^{-1}(q,\omega')
  A_a(\vec p-\vec q,\omega)(1+n_B(\omega')-f_a(\omega))}{
  z-\omega'-\omega}\;\;\;\;,
\end{eqnarray}
whereas the corresponding quasi-particle expression is given by
\begin{eqnarray}
  \label{qt_self}
  \Sigma_a^{corr,QT}(p,z) & = &
  -\!\int_{-\infty}^{\infty}\!\frac{d\omega'}{(2\pi)}\!
  \int\!\frac{d^3q}{(2\pi)^3}
  V(q)\frac{2\,\mbox{Im}\,\epsilon^{-1}(q,\omega')
  (1+n_B(\omega')-f_a(E_a(\vec p-\vec q)))}{
  z-\omega'-E_a(\vec p-\vec q)}\;\;\;\;,
\end{eqnarray} 
Here, $V(q)$ denotes the Coulomb potential, $n_B(\omega')\,=\,
\left(\rm exp(\beta\,\omega')-1 \right)^{-1}$ and $f_a$ the Fermi
function of the specie $a$. The solutions of
\begin{eqnarray}
  \label{quasi-particle}
  E_a(p) & = & \frac{\hbar^2\,p^2}{2\,m_a} + \mbox{Re}\,
 \Sigma_a(p,E_a(p))
\end{eqnarray}
defines the quasi-particle energy.
 Besides the spectral function $A_a$, medium
modifications enter via the dielectric function $\epsilon(q,\omega)$.
The set of equations
(\ref{dyson}) and (\ref{hedin}) are to be solved self-consistently.
Furthermore, the dielectric function depends on the spectral function
as well. Here, the RPA expression for the dielectric function is used,
i.e. self-energy effects as well as vertex correction in the
polarization function are ignored. The consistency of this
approximation will be discussed below. In the classical limit the
dielectric response function can be calculated analytically, yielding
\cite{kke86}:
\begin{equation}
  \epsilon(q,\omega)=1+\sum_{c=1}^3\,\frac{\kappa_c^2}{q^2}\,
  \left[1-2\,x_c\,\mbox{exp}(-x_c^2)\,\int_0^{x_c}\!dt \, 
  \mbox{exp}(t^2)\;+i\sqrt{\pi}\,x_c\,\mbox{exp}(-x_c^2)
\right]
\end{equation}
with the abbreviations
$\kappa_c=\sqrt{\frac{Z_c^2\,e^2\,n_c}{\epsilon_0\,k_B\,T}}$ and 
$x_c=\frac{\omega}{q}\,\sqrt{\frac{m_c}{2\,k_B\,T}}$.
Additional sum rules hold (see \cite{bh96})
\begin{eqnarray}
  \int_{-\infty}^{\infty}\!d\omega\,\omega\,A_a(p,\omega) & = &
  E^{HF}(p) \;\;\;,\\
  \int_{-\infty}^{\infty}\!d\omega\,\omega^2\,A_a(p,\omega) & = &
  \int_{-\infty}^{\infty}\!d\omega\mbox{Im}\,\Sigma_a(p,\omega)\,+\,
  \left(E^{HF}(p)\right)^2 \;\;\;,
\end{eqnarray}
which present a convenient check of the numerics involved in the
self-consistent solution.  $E^{HF}(p)\,=\,\hbar^2\,p^2/(2\,m)\,+\,
\Sigma^{HF}(p)$ is the quasi-particle energy in
Hartree-Fock approximation.

\section{Self-consistent determination of the spectral function
         within the GW approximation}
\label{self}

\subsection{Results for the spectral function}

Using the RPA expression as an input, the spectral function can be
determined by solving the set of equations (\ref{dyson}) and (\ref{hedin})
iteratively  until stability is reached. To start the iteration, one
can use a quasi-particle picture or a lorentzian approximation of the
spectral function with a certain width. This width can be used to
accelerate the convergence of the iteration method. The
self-consistently determined spectral function is given in figure
1 for the solar core plasma. In accordance with solar models
\cite{bp95}, the temperature is assumed to be 
$T_{\odot}\,=\,15.6\,\times\,10^6\,\mbox{K}$ and the density $ n\,=\,156\,
g/cm^3$. The plasma consists of three components, electrons, protons
and alpha-particles, with a hydrogen mass fraction of 33\%.
The spectral function is shown as a function of the frequency for a
fixed momentum. The spectral function is fairly broad, its width is
about a fifth of the thermal energy. Since the function is asymmetric,
the definition of a width and a shift is to a certain extend
ambiguous. The definition of an effective quasi-particle description
will be proposed in section \ref{effective}. An undamped quasi-particle is
by no means an adequate description of the spectral function. Therefore,
the calculation of thermodynamical properties should be based on the 
spectral function calculated here, instead of a quasi-particle
approach. In figure 2, a contour plot of the spectral
function as a function of the energy and the momentum is shown. 
The plot shows that the situation discussed above applies also to
higher momenta. It has been found that the above given sum rules are 
fulfilled within the numerical accuracy.

In figure 3, the self-consistently determined energy is
shown as a function of the frequency along with the quasi-particle
self-energy.  As found earlier \cite{nb86,fkr95} the
quasi-particle self-energy shows a logarithmic singularity at the
plasmon energy. Due to the additional integration in the definition of
the self-energy the self-consistent one does not exhibit a singularity
anymore.  This corresponds to results reported by Alm et al.
\cite{ars96}, where the spectral function and the self-energy are
calculated self-consistently for nuclear matter. Since the forces in
nuclear matter are short-ranged, the important feature discussed there
is the formation of bound states. This issue is of lower relevance for
the high temperatures considered here.

\subsection{Effective quasi-particle picture}
\label{effective}

For the sake of comparison, we define an effective quasi-particle
by fitting the self-consistently determined spectral function to a
lorentzian shape. This can be achieved by solving the dispersion
relation
\begin{eqnarray}
  \label{dispersion}
  \omega-\frac{\hbar^2\,p^2}{2\,m_e}-\Sigma^F_e(p) & = &
  \Sigma^{\rm corr}_e(p,\omega)\;\;\;.
\end{eqnarray}
Here, $\Sigma^F_e$ denotes the Fock shift.
The solution $\omega_0$ of this equation can be regarded as the 
quasi-particle energy. $\Delta(p)^{\rm eff}\,=\,\mbox{Re}\,\Sigma(p,\omega_0)$ 
and $\Gamma^{\rm eff}(p)\,=\,\mbox{Im}\,\Sigma(p,\omega_0)$ are interpreted as
the shift and the width of a finite lifetime quasi-particle.
However, a lorentzian fit is only a crude approximation to the
self-consistent spectral function, e.g. the thermal average of the
effective quasi-particle shift is not exactly the Debye
shift $\kappa\,e^2/2$ contrary to the quasi-particle result \cite{kke86}. 
A comparison of the effective
quasi-particle shift and the quasi-particle self-energy based on a free
dispersion is given  in figure 4. Note, that the steep
decrease of the quasi-particle self-energy at small momenta is not
found in the effective shift, while the high momentum behaviour is
almost identical. Therefore, the shift is overestimated in a simple
quasi-particle picture using a free dispersion relation. Solving the
self-consistency relation of the quasi-particle picture 
(\ref{quasi-particle}),
a shift corresponding to the one reported here was found by
Fehr et al. \cite{fkr95}.

Using the effective shift and width defined above, the temperature and
density dependence of the spectral function can be studied. The
results are shown in figure 5. The width is given as a
function of the density for different temperatures. At small as well
as at high densities, the width decreases, showing a restoration of
the quasi-particle picture. The maximum in between is shifted to
higher densities with increasing temperature. Furthermore, comparing
different temperatures at a fixed density, a strong temperature
dependence is observed. This implies, that thermal collisions are the
driving mechanism behind the broadening a the spectral function.

\subsection{Implications for thermodynamical properties}

In a first step, the implications for the electron chemical potential
due to the improved determination of the spectral function will be
examined.  This has important consequences for the so called Salpeter
correction \cite{sal54} to thermonuclear reaction rates.  Using the
self-consistent spectral function the chemical potential for the solar
core plasma is determined from equation (\ref{density_relation}) to be
-146.5 Ryd, whereas the chemical potential of an ideal electron gas
would be -142.9 Ryd. Using a quasi-particle approximation it is -146.3
Ryd, showing that the broadening of the spectral function has little
influence on the chemical potential.  Using the Salpeter correction,
i.e considering a Debye-shift only, the chemical potential results in
-146.2 Ryd.  Therefore, on the level of single particle corrections,
the reaction rates can be excellently described by the Salpeter
expression.  Nevertheless, the dynamics also enter in two-particle
corrections, which will be considered in a forthcoming paper
\cite{wr97}.  Besides the chemical potential, the effects on the
equation of state can be studied.

\section{Conclusions}
\label{conclude}

The self-consistent determination of the spectral function within the
screened potential approximation is reported. For the solar core
plasma, the self-consistent spectral function is found to be fairly
broad.  The quasi-particle picture is not an adequate description of
the solar core plasma. Therefore, the calculation of thermodynamic
properties of the solar core plasma should be based on the spectral
function given above. A lorentzian approximation of the spectral
function is defined introducing an effective shift and width. This can
be interpreted as a damped quasi-particle description. It has been
found, that a quasi-particle description overestimates the self-energy
at small momenta to a large extend. Furthermore, the
self-consistently determined energy does not exhibit a logarithmic
singularity at the plasmon energy contrary to the quasi-particle one.

It has been pointed out, that the scheme given above is not completely
self-consistent, since the Green's function in the RPA bubble is not
iterated. However, a complete iteration \cite{gbh95} shows that the
dielectric function does not obey exactly known properties like sum
rules. This is due to the fact, that besides self-energy corrections
also vertex corrections are to be included. Within the Green's
function method, Baym and Kadanoff \cite{bk61} developed a technique
to construct the vertex correction in a way to fulfill the Ward
identities. In this approach the sum rules are automatically obeyed.
Unfortunately, the integral equation connected with the vertex
corrections is very involved. An alternative approach, starting from
the Zubarev formalism of the non-equilibrium statistical operator has
been developed, where correlations and collisions are incorporated on
the same footing \cite{rw97}.  A compensation of self-energy and vertex
corrections to a large extend has been found, justifying the use of
the original RPA expression.

Figure Captions: \\
{\bf Fig.1} Self-consistently determined spectral function of the electrons
in the solar core plasma. The momentum $p$ is fixed to $p=0.21
\frac{1}{a_B}$. The self-consistent result as well as 
the first iteration step starting from a Lorentzian
initialisation of the spectral function are shown. \\
{\bf Fig.2} Contour plot of the self-consistently determined spectral function
  of the electrons in the solar core plasma. The spectral function is
  given as a function of frequency and momentum. Note that the
  spectral function remains fairly broad at higher momenta. \\
{\bf Fig.3} The quasi-particle self-energy and the
self-consistently determined self-energy as a function of the
frequency at a fixed momentum. Note the logarithmic singularity
of the imaginary part of the self-energy at the plasma frequency
$\omega_{pl}=\pm21.15\,\mbox{Ryd}$. \\
{\bf Fig.4} The effective quasi-particle shift as a function of the
  wave number in comparison with the quasi-particle shift using a free
  dispersion relation.\\
{\bf Fig.5} The effective width of the spectral function as a function of the
density in the long wavelength limit. The temperature is used as a parameter.
\\

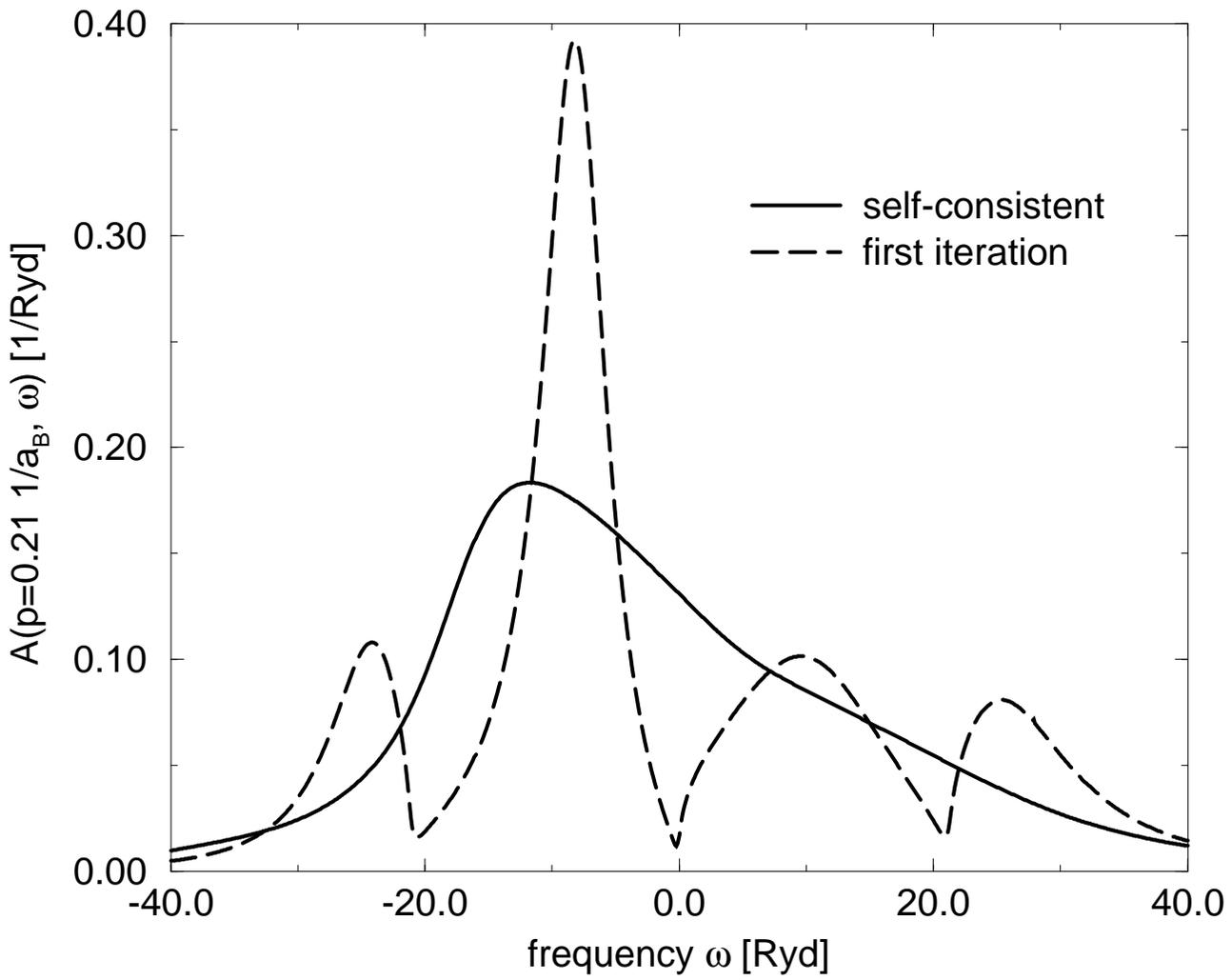
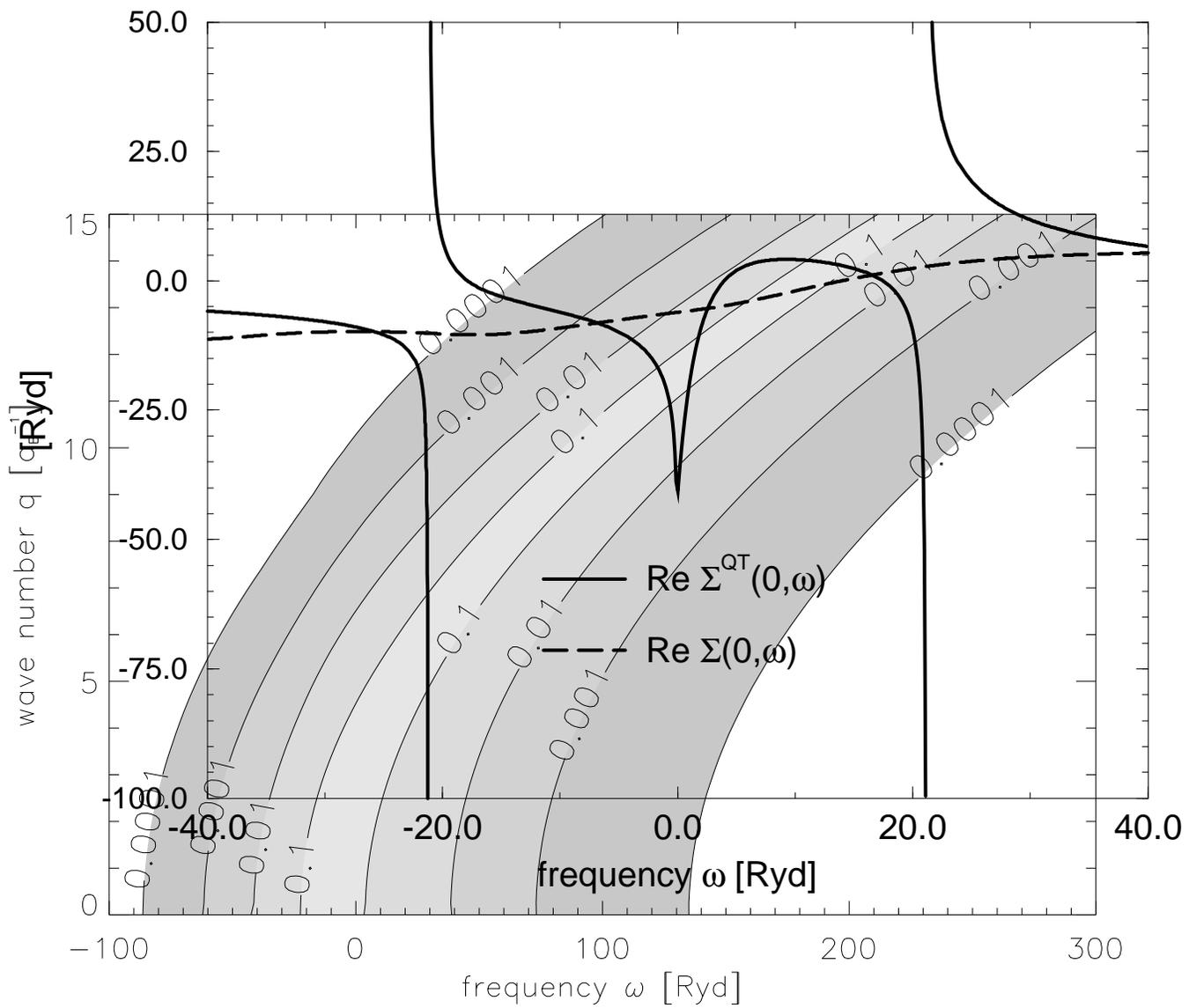
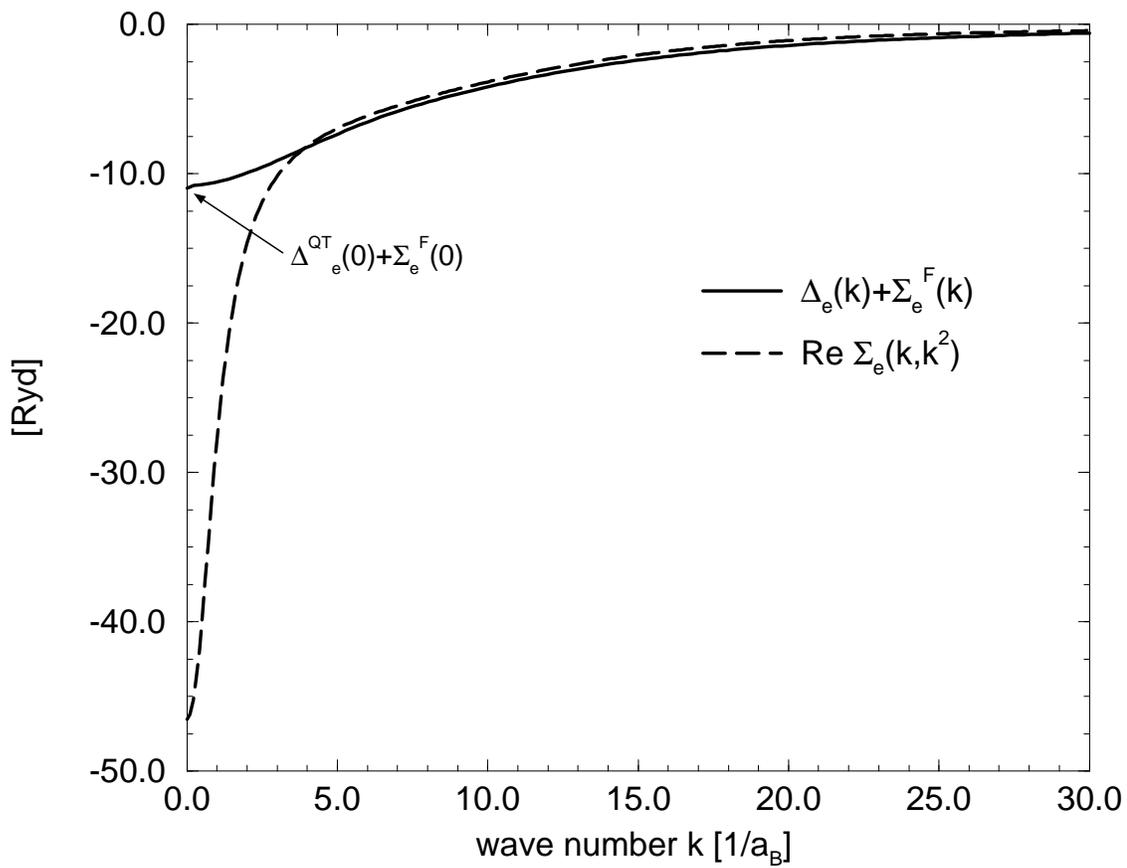
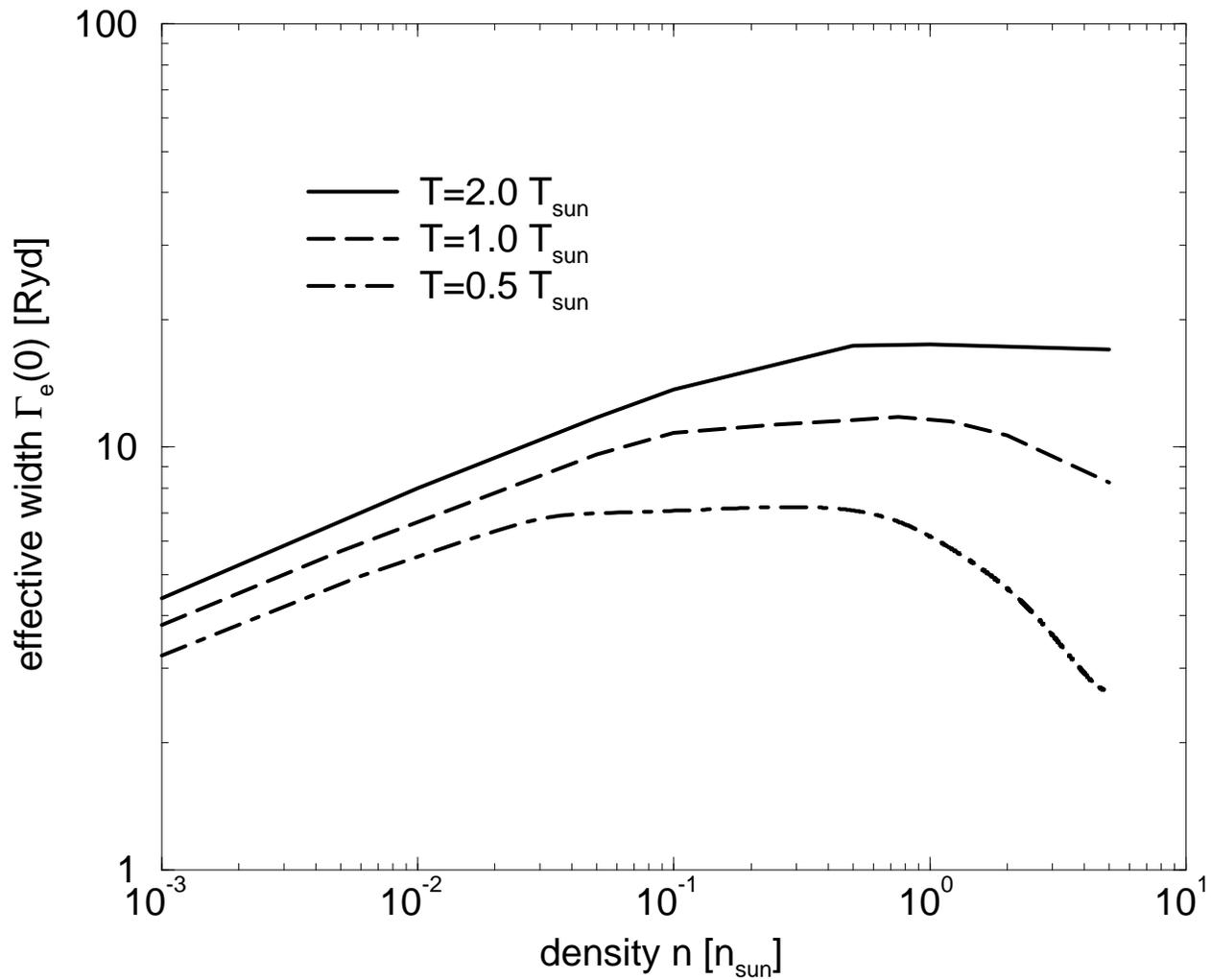
\end{document}